\documentclass[a4paper,11pt]{article}
\usepackage{jheppub} % for details on the use of the package, please
\usepackage[T1]{fontenc} % if needed

\usepackage{relsize}
\usepackage{mathtools}
\usepackage{empheq}

\title{\boldmath $T\bar{T}$ deformation of classical Liouville field theory}

\author[a]{Matias Leoni}
\affiliation[a]{Physics Department, University of Buenos Aires FCEN-UBA and IFIBA-CONICET\\Ciudad Universitaria, pabell\'on 1, 1428, Buenos Aires, Argentina.}
\emailAdd{leoni@df.uba.ar}
\def\vp{\varphi}
\def\dvp{\partial\varphi}
\def\dbvp{\bar\partial\varphi}
\def\d{\partial}
\def\db{\bar\partial}
\def\t{\mathlarger{\mathlarger{\tau}}}

\abstract{We consider the irrelevant flow of classical Liouville field theory driven by the $T\bar T$ operator. After discussing properties of its exact action and equation of motion we construct an infinite set of conserved currents. We also find its vacuum solutions.}

\begin{document} 
\maketitle
\flushbottom

\section{Introduction}

Recent years have seen a lot of attention in the study of specific types of irrelevant deformations of classical and quantum field theories. Particularly, the class of deformations with Zamolodchikov's $T\bar T$ operator \cite{Zamolodchikov:2004ce,Smirnov:2016lqw,Cavaglia:2016oda} are of interest in the context of $AdS/CFT$ \cite{McGough:2016lol,Turiaci:2017zwd,Giveon:2017nie,Giveon:2017myj,Asrat:2017tzd,Giribet:2017imm,Kraus:2018xrn,Cottrell:2018skz,Baggio:2018gct,Babaro:2018cmq,Park:2018snf,Nakayama:2019mvq,Apolo:2019yfj,Giribet:2020kde,He:2019vzf} and effective String Theory\cite{Dubovsky:2012wk,Dubovsky:2012sh,Caselle:2013dra,Chen:2018keo,EliasMiro:2019kyf}. A deformation by a given operator of a known field theory induces a flow which in the case of irrelevant deformations is driven by the deformation at high energies. This in general means that they are generically harder to study as compared to relevant deformations which drive the flow in the opposite regime. One of the interesting features of $T\bar T$ deformations is that not only do they have a flow which can be determined in many cases but also the flow seems to preserve integrable structures. These type of irrelevant deformations are particularly interesting in two dimensional field theories where many known examples exist of classical and quantum integrability without the need to resort to supersymmetry. 

A particular class of two dimensional field theories where the $T\bar T$ flow can be followed exactly are scalar field theories with a background independent potential. The closed form of their Lagrangian was first obtained in \cite{Bonelli:2018kik,Tateo:2017igst} and they constitute a rich ground where integrability under the $T\bar T$ deformation can be studied. In fact the authors of \cite{Conti:2018jho} were able to construct the $T\bar T$-deformed Lax Pair of the sin(h)-Gordon model thus revealing a nice geometrical interpretation of the $T\bar T$ deformation \cite{Conti:2018tca,Aharony:2018bad} (see also \cite{Dubovsky:2017cnj,Dubovsky:2018bmo,LeFloch:2019rut,Conti:2019dxg}).

Classical Liouville Field theory (LFT) has been studied for more than a hundred years \cite{Liouville:1853,Poincare:1893}. It is a field theory which describes the conformal factor of a two-dimensional space of constant curvature and its relation to the classical uniformization problem provides interesting connections between field-theory and two dimensional geometry \cite{Takhtajan:1994vt,Cantini:2001wr,Cantini:2002jw,Hadasz:2003kp}. In more recent decades its full quantum version has been bootstrapped \cite{Dorn:1994xn,Zamolodchikov:1995aa} and it is an important piece in the worldsheet formulation of String theories and two dimensional gravity theories \cite{Giribet:2001ft,Ribault:2005wp,Hikida:2007tq}. In connection to its integrability \cite{DHoker:1982wmk,Aoki:1992qu} classical LFT provides the simplest integrable equation underlying the problem of minimal surfaces embedded in $\mathbb{R}^3$.

In this work we consider classical Liouville Field theory in flat space and we study its $T\bar T$ deformation. Our motivation to study this deformation of classical LFT is to initiate the study of the deformation of one of the simplest but non-trivial conformal field theories whose integrability can be formulated within many of the usual frameworks such as $a)$ the existence of infinite integrals of motion, $b)$ Lax-pair formulation  and $c)$ B\"acklund transforms. The work is organized as follows. In the next section we begin by reviewing classical aspects of the $T\bar T$ deformation of a free scalar theory, and after explaining general results of LFT we move to section 3 where we study novel characteristics of the $T\bar T$-deformed version of LFT. After rederiving its exact Lagrangian and discussing some aspects of its equation of motion we show it is possible to construct an infinite set of higher conserved currents which generalize LFT undeformed holomorphic currents. We end the section by presenting the vacuum solutions of the $T\bar T$ deformed theory which generalize classical LFT vacua. We leave section 4 for discussion and open problems.

\section{Free scalar field and Liouville field theory}

\subsection{Free scalar field and its $T\bar{T}$ deformation}

The simplest 2-dimensional conformal field theory one can study in the context of $T\bar{T}$ deformations is the theory of a free scalar field and many of the results of this work will be generalizations of those that had been obtained for that theory. Thus, we find it useful, both for fixing notations and appreciating the generalization, to make a short review of some known results for the $T\bar{T}$ deformation of the free scalar field.

We consider a two dimensional flat space with euclidean metric and we choose the coordinates $z=x+i y$ and $\bar{z}=x-i y$. Derivatives become $\partial=\tfrac{1}{2}(\tfrac{\partial}{\partial x}-i\tfrac{\partial}{\partial y})$ and $\bar\partial=\tfrac{1}{2}(\tfrac{\partial}{\partial x}+i\tfrac{\partial}{\partial y})$. In this language the undeformed Lagrangian is
\begin{equation}
\mathcal{L}^{(0)}=\partial\varphi \bar\partial\varphi=\frac{1}{4}\partial_{\mu}\varphi\partial^{\mu}\varphi
\end{equation}
and the equations of motion are simply $\Box\varphi=4\,\partial\bar\partial\varphi=0$. An obvious conserved current of the theory is $J_{\mu}=\partial^\mu\varphi$, such that $\partial^\mu J_{\mu}=0$. In complex variables this means $\t^{(0)}_1=-\partial\varphi$ is holomorphic and $\bar{\t}^{(0)}_{-1}=-\bar\partial\varphi$ is anti-holomorphic. One can construct an infinite set of traceless symmetric products of the currents such as
\begin{align}
& J_{{\mu}_1 {\mu}_2}=J_{{\mu}_1} J_{{\mu}_2}-\tfrac{1}{2}\eta_{{\mu}_1 {\mu}_2}J^{{\rho}}J_{{\rho}}
\nonumber\\
& J_{{\mu}_1 {\mu}_2 {\mu}_3}=J_{{\mu}_1} J_{{\mu}_2}J_{{\mu}_3} - \tfrac{1}{4}\eta_{{\mu}_1 {\mu}_2}J^{{\rho}}J_{{\rho}} J_{\mu_3}- \tfrac{1}{4}\eta_{{\mu}_1 {\mu}_3}J^{{\rho}}J_{{\rho}} J_{\mu_2}- \tfrac{1}{4}\eta_{{\mu}_2 {\mu}_3}J^{{\rho}}J_{{\rho}} J_{\mu_1}
\end{align} 
and so on, which are also conserved $\partial^{\mu_1} J_{{\mu}_1... {\mu}_N}=0$. In complex variables this is translated to the simple statement that the powers $\t^{(0)}_n=-(\partial\varphi)^n$ are holomorphic (and their counterparts anti-holomorphic). The canonical stress tensor components are defined as $\t=-2 T_{zz}$, $\bar\t=-2 T_{\bar z\bar z}$ and $\Theta=2 T_{\bar{z}z}$ such that $\bar\partial\t=\partial\Theta$ and $\partial\bar\t=\bar\partial\Theta$ by Noether's theorem. In this particular case $\Theta=0$ (the canonical stress tensor is traceless) and $\t=\t^{(0)}_2$ is holomorphic $\bar\partial\t=0$ (and so on for its anti-holomorphic counterpart). 

$T\bar{T}$ deforming the theory means introducing a parameter $t$ and finding a $t$-dependent Lagrangian such that
\begin{equation}\label{TTbarDefScalar}
\frac{\partial\mathcal{L}^{(t)}}{\partial t}=\det(T_{\mu\nu}),\qquad\mbox{with}\quad \mathcal{L}^{(0)}=\partial\varphi \bar\partial\varphi
\end{equation}
and where $T_{\mu\nu}$ is the stress tensor of the theory derived from $\mathcal{L}^{(t)}$. The exact Lagrangian for $T\bar T$ deformed scalar field theory was first obtained in \cite{Cavaglia:2016oda} and it is given by
\begin{equation}\label{scalarDefLag}
\mathcal{L}^{(t)}=\frac{1}{2 t}\left(\sqrt{1+4t \partial\varphi \bar\partial\varphi}-1\right)=
\frac{1}{2 t}\left(\sqrt{1+4t X}-1\right)=X-X^2\,t+\mathcal{O}(t^2)
\end{equation}
where we shall denote $X=\partial\varphi \bar\partial\varphi$ throughout the rest of this work. Up to an irrelevant constant this action can be seen as the static gauge Nambu-Goto action of a string in a three dimensional target. The equation of motion derived from this Lagrangian is
\begin{equation}
\partial\left(\frac{\bar\partial\varphi}{\sqrt{1+4t X}}\right)+\bar\partial\left(\frac{\partial\varphi}{\sqrt{1+4t X}}\right)=0
\end{equation}
which can also be written as
\begin{equation}\label{eomScalar}
\partial\bar\partial\varphi=t\,\frac{\partial^2\varphi(\bar\partial\varphi)^2+\bar\partial^2\varphi(\partial\varphi)^2}{1+2 t X}
\end{equation}
It is worth observing that this second way of writing the equation of motion shows us that we can always exchange mixed $\partial$ and $\bar\partial$ derivatives of the field with ``pure'' derivatives $\partial,\partial^2,\bar\partial,\bar\partial^2$. While this seems a trivial observation, it continues to be true even for higher derivatives. By taking $\partial$ and $\bar\partial$ of the equation of motion (\ref{eomScalar}) we obtain a two dimensional linear system for $\partial^2\bar\partial\varphi$ and $\bar\partial^2\partial\varphi$ which can be inverted to obtain
%{\normalsize
\begin{align}\label{deomScalar}
&\partial^2\bar\partial\varphi= 
\frac{t(1+2tX)\partial^3\varphi(\bar\partial\varphi)^2+t^2 \bar\partial^3\varphi(\partial\varphi)^4}
{(1+tX)(1+3tX)}\nonumber\\
&+ \frac{t\,\partial\varphi}{2 X^2(1+t X)}\left[
\frac{(1+4tX)}{(1+2tX)}(\partial^2\varphi(\bar\partial\varphi)^2+\bar\partial^2\varphi(\partial\varphi)^2)^2
-(\partial^2\varphi(\bar\partial\varphi)^2-\bar\partial^2\varphi(\partial\varphi)^2)^2\right]
\end{align}
%}%
and an analogous expression exchanging $\partial\leftrightarrow\bar\partial$ for $\bar\partial^2\partial\varphi$. In the right hand side of the last equation we see only pure derivatives  $\partial,\partial^2,\partial^3,\bar\partial,\bar\partial^2,\bar\partial^3$. This pattern continues indefinitely and the lesson is that thanks to the structural form of the equation of motion (\ref{eomScalar}) mixed derivatives of the field can always be exchanged through the equation of motion and properties such as (\ref{deomScalar}) to pure derivatives of the field. This fact will be useful when we construct conserved currents for the $T\bar T$ deformation of LFT.

It is possible to construct an infinite set \cite{Cavaglia:2016oda} of conserved currents $\t_n$ and $\bar \t_{-n}$ for $n\geq 1$ which generalize $\t^{(0)}_n$ and $\bar \t^{(0)}_{-n}$  that satisfy
\begin{equation}\label{ConservationScalar}
\bar\partial \t_n-\partial\Theta_{n-2}=0,\qquad \partial \bar \t_{-n}-\bar\partial\bar\Theta_{2-n}=0
\end{equation}
from which an infinite set of local integrals of motion can be written down
\begin{equation}
Q_{n-1}=\oint (\t_n\, dz+\Theta_{n-2}\, d\bar z),\quad 
\bar Q_{1-n}=\oint (\bar \t_{-n}\, d\bar z+\bar \Theta_{2-n}\, dz)
\end{equation}
The higher conserved currents are
\begin{align}\label{HigherCurrentsScalar}
& \t_n=-\frac{(\partial\varphi)^n}{\sqrt{1+4tX}}\left(\frac{2}{1+\sqrt{1+4tX}}\right)^{n-2}=
-(\partial\varphi)^n\left(1-n X t+\mathcal{O}(t^2)\right)\nonumber\\
& \Theta_{n-2}=-t\frac{(\partial\varphi)^n(\bar\partial\varphi)^2}{\sqrt{1+4tX}}\left(\frac{2}{1+\sqrt{1+4tX}}\right)^{n}=
-t X^2(\partial\varphi)^{n-2}+\mathcal{O}(t^2)
\end{align}
and the property (\ref{ConservationScalar}) can be verified with the use of the equations of motion (\ref{deomScalar}). Similar expressions with $\partial\leftrightarrow\bar\partial$ hold for $\bar \t_{-n}$ and $\bar\Theta_{2-n}$. These set of higher conserved currents include the components of the stress tensor
\begin{align}\label{tauthetascalar}
&\t_2=-\frac{(\partial\varphi)^2}{\sqrt{1+4tX}}=-\mathcal{L}^{(t)}_X (\partial\varphi)^2=\t\nonumber\\
&\Theta_0=-\frac{(\sqrt{1+4tX}-1)^2}{4t\sqrt{1+4tX}}=X \mathcal{L}^{(t)}_X-\mathcal{L}^{(t)}=\Theta
\end{align}
where $\mathcal{L}^{(t)}_X=\partial_X\mathcal{L}^{(t)}$. 

An interesting property of Lagrangian (\ref{scalarDefLag}) is its scaling as a function of $t$ and $X$: 
\begin{equation}\label{scalingDefScalar}
\mathcal{L}^{(\lambda t)}(\lambda^{-1} X)=\lambda^{-1}\mathcal{L}^{(t)}( X)
\end{equation}
This implies
\begin{equation}
t\,\frac{\d \mathcal{L}^{(t)}}{\d t}=X \mathcal{L}^{(t)}_X-\mathcal{L}^{(t)}
\end{equation}
Notice the r.h.s. of the previous equation is the definition of $\Theta$ in (\ref{tauthetascalar}). On the l.h.s on the other hand we have, by the definition of $T\bar T$ deformation (\ref{TTbarDefScalar}), $t\,\frac{\d \mathcal{L}^{(t)}}{\d t}=t\,\mathrm{Tr(T_{\mu\nu})}=t(\Theta^2-\t\bar\t)$. Thus the operator $T\bar T(z,\bar z)=\t\bar\t-\Theta^2$ satisfies
\begin{equation}\label{noteworthy1}
t\,T\bar T(z,\bar z)=-\Theta(z,\bar z)
\end{equation}
This noteworthy property was found in \cite{Cavaglia:2016oda}, it is a consequence of the scaling properties of the Lagrangian (\ref{scalarDefLag}); see \cite{Conti:2018jho} for other important properties of these type of theories related to their scaling. The last equation implies
\begin{equation}\label{noteworthy2}
T\bar T(z,\bar z)=\frac{1}{2t^2}(\sqrt{1+4t^2\t\bar\t})
\end{equation}

\subsection{Classical Liouville field theory}
\label{LFT}

Local properties of classical Liouville field theory can be derived from the Lagrangian
\begin{equation}\label{UndefLFT}
\mathcal{L}^{(0)}=\dvp\dbvp-\mu e^{\vp}=X+V
\end{equation}
where we shall use $V=-\mu e^\vp$ for the rest of this work. The equation of motion
\begin{equation}\label{eomUndefLFT}
2\partial\dbvp+\mu e^\vp=0
\end{equation}
describes the conformal factor of a two-dimensional constant ($2\mu$) curvature space with $\mu$ having dimensions of $(Length)^{-2}$. The theory in the Riemann sphere is globally defined with the boundary condition
\begin{equation}
\vp(z,\bar z)=-2\log(z\bar z)+\mathcal{O}(1),\quad\mbox{for}\quad |z|\to\infty
\end{equation}
This boundary condition is consistent with the transformation rule for the field under holomorphic mappings $z=z(w)$
\begin{equation}\label{transformationLFT}
\vp(w,\bar w)=\vp(z,\bar z)+\log(z'(w)\bar z'(\bar w))
\end{equation}
which leave the equation of motion (\ref{eomUndefLFT}) invariant.

The canonical stress tensor derived from (\ref{UndefLFT}) is
\begin{equation}
T^{c}_{\mu\nu}=\tfrac{1}{2}\partial_\mu\vp\partial_\nu\vp
-\tfrac{1}{4}\eta_{\mu\nu}\partial_\rho\vp\partial^\rho\vp+\eta_{\mu\nu}\mu e^\vp
\end{equation}
or in complex components
\begin{equation}
\t=-(\dvp)^2,\quad \bar\t=-(\dbvp)^2,\quad \Theta=\mu e^\vp
\end{equation}
with the conservation reading $\db \t=\d\Theta$ and $\d \bar\t=\db\Theta$. We therefore see that the canonical stress tensor is not traceless and does not automatically provide an (anti)-holomorphic ($\bar\t$)$\t$. 

One can always modify the stress tensor with a total derivative $T_{\mu\nu}=T^{c}_{\mu\nu}+\partial^{\rho}B_{\rho\mu\nu}$, with $B$ being antisymmetric in its first two indexes $B_{\rho\mu\nu}=-B_{\mu\rho\nu}$ to guarantee conservation. We may choose
\begin{equation}
B_{\rho\mu\nu}=\eta_{\mu\nu}\,\d_\rho\vp-\eta_{\rho\nu}\,\d_\mu\vp
\end{equation}
With this choice the complex components of the modified stress tensor become
\begin{equation}
\t_2=-(\dvp)^2+2\d^2\vp,\quad \bar\t_{-2}=-(\dbvp)^2+2\db^2\vp, \quad \Theta_0=\mu e^\vp+2\d\db\vp=0
\end{equation}
Notice that due to the equation of motion (\ref{eomUndefLFT}) we have $\Theta_0=0$, this is, the new stress tensor is traceless which means
\begin{equation}
\db\t_{2}=0,\quad \d\bar\t_{-2}=0
\end{equation} 
{\it i.e.} the stress tensor ($\bar\t_{\!-\!2}$) $\t_2$ is (anti)-holomorphic. This new stress tensor has the property of transforming almost homogeneously under holomorphic mappings
\begin{equation}\label{stressTransformation}
\t_2(w)=(z'(w))^2\,\t_2(z)+2\{z,w\}
\end{equation}
with $\{z,w\}$ the Schwarzian derivative. When the mapping is a global conformal transformation, the transformation is homogeneous. Having (anti)-holomorphic currents it is easy to define higher conserved currents simply by taking powers
\begin{align}
&\t_{2n}=-\tfrac{1}{4^{n-1}}\left((\dvp)^2-2\d^2\vp\right)^n=(-\tfrac{1}{4})^{n-1}(\t_2)^n,
\nonumber\\
&\bar\t_{-2n}=-\tfrac{1}{4^{n-1}}\left((\dbvp)^2-2\db^2\vp\right)^n=(-\tfrac{1}{4})^{n-1}(\bar\t_{-2})^n
\end{align}
such that $\db\t_{2n}=0$ and $\d\bar\t_{\!-2n}=0$. Those are the currents we will generalize when we $T\bar T$ deform LFT. Throughout this work the subindex of different currents will refer to the Lorentz spin of the current, where $\dvp$ has spin $+1$ and $\dbvp$ has spin $-1$ \footnote{Other authors consider the Lorentz spin in absolute value. We choose to keep track of the sign.}. To count spin one just sums the number of $\d$ and $\bar\d$ in each term such that $\mathsf{s}=\#(\d)-\#(\db)$. We will say that the current $\t_{2n}$ has spin $\mathsf{s}(\t_{2n})=2n$ because each term of the current has that spin. Fields such as $X=\dvp\dbvp$ or $\d^2\vp(\dbvp)^2$ have zero spin and we call them spinless. Keeping track of spin level will be instrumental to the fact that we will work out different operator identities valid through equations of motion. Since we will see the equation of motion and the equations derived from  it preserve spin, every field identity we aim for will have to have homogeneous spin.

To find vacuum solutions of classical LFT notice the following two identities
\begin{equation}
\left(\d^2+\tfrac{1}{4}\t_2\right)\psi(z,\bar z)=0,\qquad 
\left(\db^2+\tfrac{1}{4}\bar\t_2\right)\psi(z,\bar z)=0\qquad
\mbox{with}\quad \psi(z,\bar z)=e^{-\vp(z,\bar z)/2}
\end{equation}
For a vacuum solution that has $\t_2|_{sol}=\bar\t_2|_{sol}=0$ the $\psi$ field satisfies $\d^2\psi=0$ and $\db^2\psi=0$; this means $\psi$ is both linear in $z$ and in $\bar z$. Demanding reality of $\vp$ and with $e^{\vp}=\psi^{-2}$ one finds
\begin{equation}\label{solUndeformedLFT1}
e^{\vp(z,\bar z)}=\frac{4}{\mu\left(|a z+b|^2+|cz+d|^2\right)^2}
\end{equation}
with arbitrary complex parameters $a,b,c,d$. We have assumed $\mu>0$. Plugging this ansatz into the equation of motion (\ref{eomUndefLFT}) the parameters are constrained to satisfy $ad-bc=1$. If we make an arbitrary holomorphic mapping $z=z(w)$ and transform this solution using the rule (\ref{transformationLFT}), we will obtain another solution which shall not be a vacuum one due the the inhomogeneous term in the stress tensor transformation (\ref{stressTransformation}). The exception to this is if the transformation we choose is a global conformal transformation such that the Schwarzian derivative vanishes: for such transformations a vanishing stress tensor transforms to another vanishing stress tensor y we move through the different vacuum solutions (which only means changing the values of the parameters $a,b,c,d$).

Using the constraint $ad-bc=1$ it is possible to rewrite solution (\ref{solUndeformedLFT1}) as
\begin{equation}\label{solUndeformedLFT2}
e^{\vp(z,\bar z)}=\frac{4}{\mu\left(\frac{|z-z_0|^2}{R_0}+R_0\right)^2}
\qquad\mbox{with}\quad z_0=-\frac{b\bar{a}+d\bar{c}}{|a|^2+|c|^2},\quad
R_0=\frac{1}{|a|^2+|c|^2}
\end{equation}
In this form it is possible to appreciate that the solution is actually characterized by only three real parameters: the complex point $z_0$ and the length scale $R_0$. The field $\vp$ is centered at $z=z_0$ where it reaches a maximum of $\vp=\log\left(\frac{4}{\mu R_0^2}\right)$.

\section{$T\bar{T}$ deformation of Liouville field theory}

\subsection{Exact action and properties}\label{ActionAndProperties}

The action for the $T\bar T$ deformation of a single scalar theory with an arbitrary potential was originally written down as an undetermined series in \cite{Cavaglia:2016oda} (see also \cite{Dubovsky:2013ira}). Later, the authors of \cite{Bonelli:2018kik,Tateo:2017igst} were able to find it in a closed form. Their key observation was noticing that an instance of Burgers' differential equation was satisfied by the action. Here we find it instructive to rederive it by reordering and summing the undetermined series of \cite{Cavaglia:2016oda}. While the Burgers' equation is more elegant as a way of arriving to this action, we believe the derivation we will present could be of future reference for other $T\bar T$ deformations.

We define the $T\bar T$ deformation by
\begin{equation}\label{TTbarDeformation01}
\frac{\d\mathcal{L}^{(t)}}{\d t}=\mathrm{Tr(T_{\mu\nu})}=\Theta^2-\t\bar\t,\quad\mbox{with}
\quad  \mathcal{L}^{(0)}=X+V,
\end{equation}
where recall $X=\dvp\dbvp$ and $V=-\mu e^\vp$ (though this derivation is valid for any potential). Assuming $\mathcal{L}^{(t)}=\mathcal{L}^{(t)}(X,V)$ we have
\begin{align}\label{CanonicalTauThetaDefLFT}
&\Theta=\frac{1}{2}\left(\frac{\d\mathcal{L}^{(t)}}{\d(\dvp)}\dvp
+\frac{\d\mathcal{L}^{(t)}}{\d(\dbvp)}\dbvp-2\mathcal{L}^{(t)}\right)=
\mathcal{L}^{(t)}_X X-\mathcal{L}^{(t)}\nonumber\\
&\t=-\frac{\d\mathcal{L}^{(t)}}{\d(\dbvp)}\dvp=-\mathcal{L}^{(t)}_X (\dvp)^2,\quad
\bar\t=-\frac{\d\mathcal{L}^{(t)}}{\d(\dvp)}\dbvp=-\mathcal{L}^{(t)}_X (\dbvp)^2
\end{align}
Defining a power expansion $\mathcal{L}^{(t)}=\sum_{n=0}^{\infty}t^n\mathcal{L}_n$ and using the defining property (\ref{TTbarDeformation01}) after some elementary manipulations we obtain the recurrence
\begin{equation}\label{RecurrenceLagrangian}
\mathcal{L}_{n+1}=\frac{1}{n+1}\sum_{k=0}^n\left(
\mathcal{L}_k\, \mathcal{L}_{n-k}-2 X \mathcal{L}_{n-k}\,\mathcal{L}_{k,X}
\right)
\end{equation}
where $\mathcal{L}_{k,X}=\d_X\mathcal{L}_k$. Using this recurrence starting with $\mathcal{L}_0=X+V$ the first few terms (see appendix \ref{app:Lagrangian}) allow us to recognize the following pattern 
\begin{align}
&\mathcal{L}_n=V^{n+1}+(-1)^n\sum_{k=0}^{[\tfrac{n}{2}]}
\frac{(2n-2k)!\, X^{n-k+1}\,V^k}{k!\,\,(n-k+1)!\,(n-2k)!}\quad\mbox{and} \label{ExactSum01.1}\\
&\mathcal{L}_{n,X}=(-1)^n\sum_{k=0}^{[\tfrac{n}{2}]}
\frac{(2n-2k)!\, X^{n-k}\,V^k}{k!\,\,(n-k)!\,(n-2k)!}\label{ExactSum01.2}
\end{align}
Observe from (\ref{ExactSum01.1}) that $\mathcal{L}_n(X=0,V)=V^{n+1}$ such that $\mathcal{L}^{(t)}(X=0,V)=\tfrac{V}{1-t\,V}$. It will be easier to work out the series resulting for $\d_X\mathcal{L}^{(t)}=\sum_{t=0}^\infty t^n \mathcal{L}_{n,X}$, therefore we focus on (\ref{ExactSum01.2}). After some manipulations detailed in appendix \ref{app:Lagrangian} we can write
\begin{equation}\label{ExactSum02}
\mathcal{L}^{(t)}_X=\frac{1}{\sqrt{\pi}}\sum_{k=0}^{\infty}\sum_{n=0}^\infty\frac{\Gamma(\tfrac{1}{2}+n+k)}
{k!\,n!}(-4tX)^n(4t^2XV)^k
\end{equation}
But now this double series became an iterated double binomial series which can be easily summed with the usual formulae 
\begin{equation}
\mathcal{L}^{(t)}_X=\frac{1}{\sqrt{1+4tX(1-tV)}}
\end{equation}
Integrating this in $X$ and using the previous result $\mathcal{L}^{(t)}(X=0,V)=\tfrac{V}{1-t\,V}$ we finally get
\begin{equation}\label{DefLFTLag1}
\mathcal{L}^{(t)}(X,V)=\frac{1}{2t(1-tV)}\left(\sqrt{1+4tX(1-tV)}+2tV-1\right)
\end{equation}
or, more explicitly for our potential of interest 
\begin{equation}\label{DefLFTLag2}
\mathcal{L}^{(t)}=\frac{1}{2t(1+t\mu e^\vp)}\left(\sqrt{1+4t\dvp\dbvp(1+t\mu e^\vp)}-2t\mu e^{\vp}-1\right)
\end{equation}
This is the $T\bar T$ deformed action of classical LFT which is the main interest of our work. It is an irrelevant deformation of LFT with a deforming flow defined through the determinant of its canonical stress tensor. In section \ref{sec:discussion} we shall discuss possible alternatives of this type of deformation.

The analogous scaling property we saw in the $V=0$ case is now ({\it c.f.} (\ref{scalingDefScalar}))
\begin{equation}\label{scalingDefLFT}
\mathcal{L}^{(\lambda t)}(\lambda^{-1}X,\lambda^{-1}V)=\lambda^{-1}\mathcal{L}^{(t)}(X,V)
\end{equation}
which implies
\begin{equation}
t\frac{\d\mathcal{L}^{(t)}}{\d t}=X\mathcal{L}^{(t)}_X-\mathcal{L}^{(t)}+V\mathcal{L}^{(t)}_V
\end{equation}
In our case $V\mathcal{L}^{(t)}_V=\frac{\d \mathcal{L}^{(t)}}{\d\vp}$. With $\Theta$ from (\ref{CanonicalTauThetaDefLFT}) we find the analogous to (\ref{noteworthy1})
\begin{equation}
t\,T\bar T(z,\bar z)=-\Theta(z,\bar z)-\frac{\d \mathcal{L}^{(t)}}{\d\vp}
\end{equation} 
which also implies
\begin{equation}
T\bar T(z,\bar z)=\frac{1}{2t^2}\left(
\sqrt{1+4t^2\t\bar\t+4t\,\frac{\d \mathcal{L}^{(t)}}{\d\vp}}-1-2t\,\frac{\d \mathcal{L}^{(t)}}{\d\vp}
\right)
\end{equation}

The equation of motion derived from (\ref{DefLFTLag2}) is
\begin{equation}\label{eomTTLFT1}
\d\left(\frac{\dbvp}{\Omega}\right)+\db\left(\frac{\dvp}{\Omega}\right)=
\frac{V\left(1+\Omega\right)^2}{4 \Omega(1-tV)^2}
\end{equation}
where for the rest of this work we will use 
\begin{equation}
\Omega\equiv\Omega(X,V)=\sqrt{1+4tX(1-tV)}=\sqrt{1+4t\dvp\dbvp(1+t\mu e^\vp)}
\end{equation}
Expanding the derivatives we arrive to the following form of the equation of motion
\begin{align}\label{eomTTLFT2}
\d\dbvp=&\frac{V (1+\Omega )^2 (2 \Omega-1)}{4 (1-t V)^2 \left(1+\Omega ^2\right)}+
\frac{2 t (1-t V)}{1+\Omega ^2}\Big(\d^2\vp(\dbvp)^2+\bar\d^2\vp(\dvp)^2\Big)\nonumber\\
=&\frac{V}{2}+t\Big(V^2+2VX+\d^2\vp(\dbvp)^2+\bar\d^2\vp(\dvp)^2\Big)+\mathcal{O}(t^2)
\end{align}
Here we once again see that the equation of motion allows us to replace mixed derivatives $\d\dbvp$ with pure derivatives $\dvp,\dbvp,\d^2\vp,\db^2\vp$. Moreover, continuing this pattern, by taking $\d$ and $\db$ derivatives of the equation of motion (\ref{eomTTLFT2}) we get a 2 dimensional linear system on $\d^2\dbvp$ and $\db^2\dvp$ which can easily be inverted. The solution will allow us to express triple mixed derivatives $\d^2\dbvp$ and $\db^2\dvp$ in terms of pure derivatives $\dvp,\dbvp,\d^2\vp,\db^2\vp,\d^3\vp,\db^3\vp$. This pattern continues {\it ad inf.} and it is a useful property to decide how to construct the higher conserved currents of the theory. It should be observed that both the equation of motion (\ref{eomTTLFT2}) and the expressions for $\d^2\bar\dvp$ and $\bar\d^2\dbvp$ preserve the {\it spin} as defined in the previous section.

While by deforming LFT we have broken conformal symmetry, the equation of motion (\ref{eomTTLFT2}) is still covariant under the complex transformation $z=a z'+b$, $\bar{z}=\bar a\bar z'+\bar b$ if the field and the $t$ parameter transform as
\begin{equation}\label{scalingSymmetry}
\vp(z',\bar z')=\vp(z,\bar z)+\log|a|^2,\qquad t'=|a|^{-2}t
\end{equation}
as can be easily checked explicitly in (\ref{eomTTLFT2}). This symmetry, akin of scaling (\ref{scalingDefLFT}), will be useful when we study vacuum solutions of the theory.

Finally, the explicit form of the components of the canonical stress tensor are
\begin{equation}\label{CanonicalTauThetaDefLFT2}
\t=-\frac{1}{\Omega}(\dvp)^2,\quad
\bar\t=-\frac{1}{\Omega}(\dbvp)^2\quad
\Theta=-\frac{(1-\Omega)^2}{4t\Omega(1-tV)}-\frac{V}{1-tV}
\end{equation}
such that they satisfy $\bar\d\t-\d\Theta=0$ and $\d\bar\t-\bar\d\Theta=0$ by construction and can be used to compute the local integrals of motion
\begin{align}\label{ChargesCanonical}
Q=\oint(\t\,dz+\Theta\,d\bar z),\qquad
\bar Q=\oint(\bar\t\,d\bar z+\Theta\,d z)
\end{align}

\subsection{Higher Conserved currents}

We would like to show it is possible to generalize the undeformed currents of LFT we presented in section \ref{LFT}
\begin{equation}\label{undefCurrents}
\t^{(0)}_{2n}=-\tfrac{1}{4^{n-1}}\left((\dvp)^2-2\d^2\vp\right)^n,\quad \bar\t^{(0)}_{-2n}=-\tfrac{1}{4^{n-1}}\left((\dbvp)^2-2\db^2\vp\right)^n
\end{equation}
to our $T\bar T$ deformed theory. Specifically we would like two sets $\{\t_{2n},\Theta_{2n-2}\}$ and $\{\bar\t_{-2n},\bar\Theta_{-2n+2}\}$ with $n\geq 1$ such that
\begin{equation}\label{DefConservation}
\bar\d\t_{2n}-\d\Theta_{2n-2}=0, \qquad
\d\bar\t_{-2n}-\bar\d\bar\Theta_{-2n+2}=0
\end{equation}
From these one could construct the set of charges
\begin{equation}\label{DefCharges}
Q_{2n\!-\!1}=\oint (\t_{2n}\, dz+\Theta_{2n\!-\!2}\, d\bar z),\quad 
\bar Q_{1\!-\!2n}=\oint (\bar \t_{\!-\!2n}\, d\bar z+\bar \Theta_{2\!-\!2n}\, dz)
\end{equation}
We expect $\t_{2n}\to\t^{(0)}_{2n}$ and $\Theta_{2n-2}\to 0$ when $t\to 0$. Let us explain the strategy we followed. We begin by noticing we can write
\begin{equation}
\t^{(0)}_{2n}=-\tfrac{1}{4^{n-1}}(\dvp)^{2n}\left(1-2\frac{\d^2\vp(\dbvp)^2}{X^2}\right)^n
\end{equation}
and the simple observation that the parenthesis in the last equation is spinless. With this last fact in mind we point out that the $2n$- and $(2n\!-\!2)$-spin of  $\t_{2n}$ and $\Theta_{2n-2}$ can be set by an overall power of the first derivative of the field without loosing generality. In the function multiplying that overall power we should expect second derivatives of the field from the limit form (\ref{undefCurrents}). Moreover, as explained before, we should only use ``pure'' derivatives since mixed ones can always be replaced by the equation of motion and its derivatives. Without losing generality we considered the following two combinations containing second derivatives which come from metric contractions which avoid mixed derivatives and are spinless
\begin{equation}
Y_s=\d^2\vp(\dbvp)^2+\bar\d^2\vp(\dvp)^2,\qquad 
Y_a=\d^2\vp(\dbvp)^2-\bar\d^2\vp(\dvp)^2
\end{equation}
Notice that the combination of second derivatives of both types can be written as $\d^2\vp\bar\d^2\vp=\tfrac{Y_s^2-Y_a^2}{4X^2}$. With all these considerations our general ansatz was
\begin{align}
&\t_{2n}=-f_n(X,V,Y_s,Y_a)(\dvp)^{2n}\qquad &\Theta_{2n-2}=-g_n(X,V,Y_s,Y_a)(\dvp)^{2n-2}\nonumber\\
&\bar\t_{\!-2n}=-f_n(X,V,Y_s,\!-Y_a)(\dbvp)^{2n}\qquad &\bar\Theta_{\!-2n+2}=-g_n(X,V,Y_s,\!-Y_a)(\dbvp)^{2n-2}
\end{align}
this is, we admit a dependence on the field through $V=-\mu e^\vp$, on its first derivatives through the spin power $(\dvp)^{2n}$ and the spinless combination $X=\dvp\dbvp$, and on its second derivatives through the spinless variables $Y_s$ and $Y_a$. Plugging these in the conservation equation (\ref{DefConservation}) we use the equation of motion and its derivatives and we obtain a linear combination of powers of the higher derivatives up to $\d^3\vp$ and $\bar\d^3\vp$ whose coefficients we set to zero. Doing so we obtain a complicated system of linear first order differential equations for the functions $f$ and $g$ on its four variables. We shall not present these equations here (they are not very illuminating) but just comment that the first progress one makes in the process of solving them is to notice the variables $Y_s$ and $Y_a$ are constrained by the equations to appear in the combinations
\begin{equation}\label{defY}
Y_{\pm}=\frac{1}{2}Y_a\pm\frac{\Omega}{1+\Omega^2}Y_s
\end{equation}
or
\begin{equation}
Y_{\pm}=\frac{(1\pm\Omega)^2}{2(1+\Omega^2)}\d^2\vp(\dbvp)^2-\frac{(1\mp\Omega)^2}{2(1+\Omega^2)}\bar\d^2\vp(\dvp)^2=
\begin{cases}
\d^2\vp(\dbvp)^2+\mathcal{O}(t^2)\\
\bar\d^2\vp(\dvp)^2+\mathcal{O}(t^2)
\end{cases}
\end{equation}
with the upper sign for $\t_{2n}$ and $\Theta_{2n-2}$ and the lower sign for $\bar\t_{\!-2n}$ and $\bar\Theta_{\!-2n+2}$. This reveals that while one could construct the currents for the undeformed theory using exclusively $\d^2\vp$ or $\bar\d^2\vp$ as in (\ref{undefCurrents}), in the deformed case a combination of both derivatives forcefully appear in the currents and they do at different order in $t$. Moreover, the currents are polynomial in the variables $Y_{\pm}$.

Besides the property of the differential equations that constrain $f$ and $g$ we have just mentioned, we will not go deeper on this path. Instead, we will present the result, establish some properties and show by induction that the conservation equation (\ref{DefConservation}) holds. The currents that satisfy (\ref{DefConservation}) are
%\begin{align}
\begin{empheq}[box=\fbox]{align}\label{TTcurrentsTau}
\t_{2n}=-\frac{(\Omega+1)^2}{\Omega(1-t\,V)}\Bigg[&
\frac{(1-t V)}{(1+\Omega)^2}\left(1+2t\frac{\,V\Omega(\Omega+2)}{1+\Omega^2}\right)(\dvp)^2
\nonumber\\
& -\frac{(1-t\,V)^2}{1+\Omega^2}\d^2\vp+
16t^2\frac{(1-t\,V)^4}{(1+\Omega^2)(1+\Omega)^4}\db^2\vp(\dvp)^4
\Bigg]^n
\end{empheq}

%\end{align}
%\begin{align}
\begin{empheq}[box=\fbox]{align}\label{TTcurrentsTheta}
\Theta_{2n-2}=-\frac{4t}{\Omega}(\dbvp)^2\Bigg[&
\frac{(1-t V)}{(1+\Omega)^2}\left(1+2t\frac{\,V\Omega(\Omega+2)}{1+\Omega^2}\right)(\dvp)^2
\nonumber\\
& -\frac{(1-t\,V)^2}{1+\Omega^2}\d^2\vp+
16t^2\frac{(1-t\,V)^4}{(1+\Omega^2)(1+\Omega)^4}\db^2\vp(\dvp)^4
\Bigg]^n
\end{empheq} 
%\end{align}
and similar expressions exchanging $\d\leftrightarrow\bar\d$ for $\bar\t_{\!-2n}$ and $\bar\Theta_{\!-2n+2}$. The following properties hold
\begin{itemize}
\item It is possible to show that the conservation of the currents $\t_{2n}$ and $\Theta_{2n-2}$ is non-trivial. By this we mean it is \underline{not} possible to find a $(2n\!-\!1)$-spin current $\rho_{2n\!-\!1}$ that satisfies
\begin{equation}\label{NoGo}
\t_{2n}\stackrel{?}{=}\d\rho_{2n\!-\!1}\quad\mbox{and}\quad  \Theta_{2n-2}\stackrel{?}{=}\bar\d\rho_{2n\!-\!1}
\end{equation}
such that conservation rule (\ref{DefConservation}) becomes a trivial statement. Here $\rho_{2n\!-\!1}$ is some function of the field and its derivatives. What we mean by this claim is that, while for a particular solution, (\ref{NoGo}) is possible locally, it is not possible to find such a $\rho_{2n\!-\!1}$ generically as a field identity. In the language of \cite{Conti:2019dxg}, the closed forms
\begin{align}\label{Forms1}
&\mathfrak{T}_{2n\!-\!1}=\t_{2n}dz+\Theta_{2n\!-\!2}d\bar z\\
&\bar{\mathfrak{T}}_{1\!-\!2n}=\bar\t_{\!-\!2n}d\bar z+\bar \Theta_{2\!-\!2n}d z
\end{align}
are not exact forms as generic field identities (while, by Poincar\'e lemma, they are locally exact for a particular solution).
\item Expanding in $t$ the expressions (\ref{TTcurrentsTau}-\ref{TTcurrentsTheta})
\begin{align}
&\t_{2n}=\t^{(0)}_{2n}
+4\t^{(0)}_{2n-2}
\left(
((2n\!+\!1)V\!-\!2nX)(\dvp)^2+2((2n\!-\!1)V\!+\!2nX)\d^2\vp
\right)\ t+\mathcal{O}(t^2)\nonumber\\
&\Theta_{2n-2}=(\dbvp)^2\ \t^{(0)}_{2n}\ t+\mathcal{O}(t^2)
\end{align}
this is, $\t_{2n}\to\t^{(0)}_{2n}$ and $\Theta_{2n-2}\to 0$ for $t\to 0$ as expected. Notice that $\bar\d^2\vp$ still does not appear until $\mathcal{O}(t^2)$. 
\item We shall use in the next section the fact that the currents can be written as
\begin{equation}
\t_{2n}=-\frac{(\dvp)^{2n}\left(F_0(X,V,Y_{+})\right)^{n}}{\Omega(1\!-\!t V)(1\!+\!\Omega)^{2n-2}}
,\quad
\Theta_{2n-2}=-\frac{4t(\dvp)^{2n}(\dbvp)^2\left(F_0(X,V,Y_{+})\right)^{n}}{\Omega(1\!+\!\Omega)^{2n}}
\nonumber
\end{equation}
\begin{equation}
\bar\t_{\!-\!2n}=\!-\!\frac{(\dbvp)^{2n}\left(F_0(X,V,\!-\!Y_{-})\right)^{n}}{\Omega(1\!-\!t V)(1\!+\!\Omega)^{2n-2}}
,\ 
\bar\Theta_{\!-\!2n+2}=\!-\!\frac{4t(\dbvp)^{2n}(\dvp)^2\left(F_0(X,V,\!-\!Y_{-})\right)^{n}}{\Omega(1\!+\!\Omega)^{2n}}
\nonumber
\end{equation}
where $F_0(X,V,\pm Y_{\pm})$ is the spinless expression
\begin{equation}\label{defF0}
F_0(X,V,\pm Y_{\pm})=(1-t V)\left(1+2t\frac{\,V\Omega(\Omega+2)}{1+\Omega^2}\right)
\mp 2\frac{(1-tV)^2}{X^2}\,Y_{\pm}
\end{equation}
with $Y_{\pm}$ defined in (\ref{defY}). This means that if for a given solution, both $F_0(X,V,Y_{+})$ and $F_0(X,V,-Y_{-})$ vanish, all the higher conserved currents $\t_{2n}$, $\Theta_{2n-2}$, $\bar\t_{\!-\!2n}$ and $\bar\Theta_{\!-\!2n+2}$ also vanish.
\item The currents depend explicitly on $V=-\mu e^\vp$ such that we can take $\mu\to 0$ to recover results valid for $T\bar{T}$ deformed scalar field theory. We obtain
\begin{align}
&T^{\mathrm{scalar}}_{2n}=-\frac{(\dvp)^{2n}}{\omega(1+\omega)^{2n-2}}
\left(
1+\frac{8t^2}{(1+2tX)}
\left(\frac{\bar\d^2\vp(\dvp)^2}{(1+\omega)^2}-\frac{\d^2\vp(\dbvp)^2}{(1-\omega)^2}\right)
\right)^n\nonumber\\
&\Theta^{\mathrm{scalar}}_{2n-2}=-\frac{4t(\dvp)^{2n}(\dbvp)^2}{\omega(1+\omega)^{2n}}
\left(
1+\frac{8t^2}{(1+2tX)}
\left(\frac{\bar\d^2\vp(\dvp)^2}{(1+\omega)^2}-\frac{\d^2\vp(\dbvp)^2}{(1-\omega)^2}\right)
\right)^n
\end{align}
where $\omega=\sqrt{1+4tX}$. These second order currents closely resemble those of (\ref{HigherCurrentsScalar}) up to the parenthesis in the r.h.s.
%\item For the general proof by induction of the conservation of these higher currents we need to prove explicitly the conservation for the lowest level, this $\bar\d\t_2-\d\Theta_0=0$ and $\d\bar\t_2-\bar\d\bar\Theta_0=0$. We leave this computation for the appendix and take it for granted in the main text. 
\item There is a degeneracy of currents only at the lowest spins. Besides $\t_2$, $\bar\t_2$, $\Theta_0$ and $\bar\Theta_0$ we have the components of the canonical stress tensor $\t$, $\bar\t$ and $\Theta$ explicitly written in (\ref{CanonicalTauThetaDefLFT2}). Thus for the lowest spin conservation, any combination will provide
\begin{align}
&\bar\d\left(\alpha_1\t_2+\alpha_2\t\right)-\d\left(\alpha_1\Theta_0+\alpha_2\Theta\right)=0
\nonumber\\
&\d\left(\beta_1\bar\t_{\!-\!2}+\beta_2\bar\t\right)-\bar\d\left(\beta_1\bar\Theta_0+\beta_2\Theta\right)=0\label{degeneracyLowest}
\end{align}
for any given $\alpha_i$, $\beta_i$. We shall expand on this point below.
\item The higher currents (\ref{TTcurrentsTau})-(\ref{TTcurrentsTheta}) satisfy recurrent relations relating them at different spin. It can be easily checked that
\begin{equation}\label{RecursionTauTheta1}
\t_{2n+2}=-\frac{\Omega(1-tV)}{(1+\Omega)^2}\t_2\t_{2n},\quad
\Theta_{2n}=-\frac{\Omega(\dvp)^2}{4t X^2}\Theta_0\Theta_{2n-2}
\end{equation}
Also, currents $\t_{2n}$ and $\Theta_{2n-2}$ are related through
\begin{equation}\label{RecursionTauTheta2}
\Theta_{2n-2}=\frac{4t(1-tV)(\dbvp)^2}{(1+\Omega)^2}\t_{2n},\quad\mbox{or}\ \ 
\t_{2n}=\frac{4t(1-tV)(\dvp)^2}{(1-\Omega)^2}\Theta_{2n-2}
\end{equation}
Analogous relations hold for $\bar\t_{\!-\!2n}$ and $\bar\Theta_{\!-\!2n+2}$ by exchanging $\d\leftrightarrow\bar\d$ in (\ref{RecursionTauTheta1})-(\ref{RecursionTauTheta2}). 
\end{itemize}

The degeneracy at the lowest spin explained at (\ref{degeneracyLowest}) can be understood by the fact that the components of the stress tensor ($\t,\bar\t,\Theta$) are related to our lowest-spin currents ($\t_2,\bar\t_{\!-\!2},\Theta_0,\bar\Theta_0$) by a derivative as follows
\begin{align}\label{relationLowSpin}
&\t_2=\t+\d\Big(\rho(X,V)\,\dvp\Big),\quad 
&\Theta_0=\Theta+\bar\d\Big(\rho(X,V)\,\dvp\Big)\nonumber\\
&\bar\t_{\!-\!2}=\bar\t+\bar\d\Big(\rho(X,V)\,\dbvp\Big),\quad 
&\bar\Theta_0=\Theta+\d\Big(\rho(X,V)\,\dbvp\Big)
\end{align}
where $\rho(X,V)=\tfrac{4(1-t V)}{\Omega+1}$. This means that the conservation equation for $\t_2$ and $\theta_0$
\begin{equation}\label{proof0}
\bar\d\t_2-\d\Theta_0=\bar\d\t+\bar\d\d\Big(\rho(X,V)\,\dvp\Big)-\d\Theta-\d\bar\d\Big(\rho(X,V)\,\dvp\Big)=\bar\d\t-\d\Theta=0
\end{equation}
holds after using Noether's theorem for the canonical stress tensor. Notice, incidentally, that relations (\ref{relationLowSpin}) also imply
\begin{align}
Q_1=\oint(\t_2\,dz+\Theta_0\,d\bar z)=
&\oint(\t\,dz+\Theta\,d\bar z)=Q\nonumber\\
\bar Q_{\!-\!1}=\oint(\bar\t_{\!-\!2}\,d\bar z+\bar\Theta_0\,dz)=
&\oint(\bar\t\,d\bar z+\Theta\,d z)=\bar Q
\end{align}
this is, the lowest-spin charges we defined in (\ref{DefCharges}) coincide with the charges computed with the canonical stress tensor (\ref{ChargesCanonical}). Once again, in the language of \cite{Conti:2019dxg}, if we define the forms associated to the stress tensor
\begin{equation}
\mathfrak{T}=\t\,dz+\Theta\,d\bar z,\qquad \bar{\mathfrak{T}}=\bar\t\,d\bar z+\bar\Theta\,dz
\end{equation}
we have that relations (\ref{relationLowSpin}) mean that $\mathfrak{T}$ and $\bar{\mathfrak{T}}$ are equal to the forms $\mathfrak{T}_1$ and $\bar{\mathfrak{T}}_{\!-\!1}$ we defined in (\ref{Forms1}) up to an exact form:
\begin{equation}
\mathfrak{T}_1=\mathfrak{T}+d\left(\rho(X,V)\dvp\right),\qquad
\bar{\mathfrak{T}}_{\!-\!1}=\bar{\mathfrak{T}}+d\left(\rho(X,V)\dbvp\right)
\end{equation} 
and therefore if the forms $\mathfrak{T}$, $\bar{\mathfrak{T}}$ are closed, the forms $\mathfrak{T}_1$, $\bar{\mathfrak{T}}_{\!-\!1}$ will also be closed: $d\mathfrak{T}_1=d\mathfrak{T}=0$ and $d\bar{\mathfrak{T}}_{\!-\!1}=d\bar{\mathfrak{T}}=0$.

Having shown the conservation (\ref{DefConservation}) for $n=1$ in (\ref{proof0}) we start the inductive reasoning. We assume equation (\ref{DefConservation}) holds for $n$ and we should prove it also holds for $n\to n+1$ as a consequence. Consider the difference $\bar\d\t_{2n+2}-\d\theta_{2n}$. We have
\begin{align}
&\bar\d\t_{2n+2}-\d\theta_{2n}=\d\left(\frac{\Omega(\dvp)^2}{4t X^2}\Theta_0\Theta_{2n-2}\right)
-\bar\d\left(\frac{\Omega(1-tV)}{(1+\Omega)^2}\t_2\t_{2n}\right)\label{proof1}\\
=&4t(1-tV)\Theta_{2n\!-\!2}\t_2\left[
\frac{(\dbvp)^2}{(1+\Omega)^2}\d\left(\frac{\Omega(\dvp)^2}{4t X^2}\right)-
\frac{(\dvp)^2}{(1-\Omega)^2}\bar\d\left(\frac{\Omega(1-tV)}{(1+\Omega)^2}\right)
\right]=0\label{proof2}
\end{align}
To go from (\ref{proof1}) to (\ref{proof2}) we have extensively used the recursive properties (\ref{RecursionTauTheta1}-\ref{RecursionTauTheta2}) and the inductive hypothesis. We leave the details in the appendix \ref{app:currents}. Having eliminated every trace of the currents from inside the bracket in (\ref{proof2}), the remaining pieces still inside $\d(..)$ and $\bar\d(..)$ depend at most on first derivatives of the field. Thus, a tedious but straightforward computation we leave to the reader shows that the bracket in (\ref{proof2}) is identically zero after the equation of motion is used. Of course, a proof for the $\bar\t_{\!-\!2n}$, $\bar\Theta_{\!-\!2n\!+\!2}$ conservation in (\ref{DefConservation}) holds analogously.

It should be noticed the conserved currents we obtained (\ref{TTcurrentsTau}-\ref{TTcurrentsTheta}) through our ansantz and differential method can also be constructed using the method proposed in \cite{Conti:2019dxg} which is based on a field dependent variable transformation \cite{Conti:2018tca,Dubovsky:2017cnj}. We have checked both derivations lead to the exact same results.

\subsection{Exact Vacuum solutions}

We define a solution to be a vacuum one if all the currents $\t_{2n}$ and $\bar\t_{\!-\!2n}$ defined in the previous section vanish when evaluated in the solution. From (\ref{RecursionTauTheta2}) we see that this implies $\Theta_{2n\!-\!2}$ and $\bar\Theta_{2\!-\!2n}$ will vanish and therefore the charges $Q_{2n\!-\!1}$ and $\bar{Q}_{1\!-\!2n}$ will also vanish ({\it c.f.} (\ref{DefCharges})). Also recall that the vanishing of $Q_1$ and $\bar Q_{-1}$ implies the vanishing of the charges (\ref{ChargesCanonical}) constructed from the canonical stress tensor. 

As we saw previously, for a solution to be a vacuum one it suffices that the spinless functions defined in (\ref{defF0}) vanish, this is $F_0(X,V,Y^+)=F_0(X,V,-Y^-)=0$. One can verify that both functions vanishing at the same time requires
\begin{equation}\label{ConditionVaccum}
Y_a=\d^2\vp(\dbvp)^2
-\bar\d^2\vp(\dvp)^2
\Big{|}_{\mathrm{sol}}=0
\end{equation}
and check that, in particular, if $\vp(z,\bar z)$ is a function of $|z-z_0|$ with arbitrary $z_0$ it satisfies the condition (\ref{ConditionVaccum}). The key to arrive to vacuum solutions in closed form for the $T\bar T$ deformed theory is to study the inverse of the solution. Consider the inverse of undeformed LFT solution (\ref{solUndeformedLFT2})
\begin{equation}\label{LFTsolutionAgain}
e^{\vp(z,\bar z)}=\frac{4}{\mu\left(\tfrac{|z-z_0|^2}{R_0}+R_0\right)^2}
\qquad\rightarrow\qquad
|z-z_0|=R_0\sqrt{\tfrac{2\, \psi}{R_0}-1}
\end{equation}
where we have set the scale $\mu=1$ in the last equation and used $\psi=e^{-\vp/2}$. Notice one can always reinstate $\mu$ with the field redefinition $\vp\to\vp+\log\mu$ and therefore we shall alternatively omit or reinstate $\mu$ when convenient. Based on (\ref{LFTsolutionAgain}) our ansatz for the inverse of the deformed solution is
\begin{equation}\label{ansatz1}
|z-z_0|=r(\psi)=R_0\sqrt{\tfrac{2\, \psi}{R_0}-1}\ \Big(1 
+t\,r_1(\psi)+t^2\,r_2(\psi)+...\Big)
\end{equation}
with $\{r_i(\psi)\}_{i\in\mathbb{N}}$ functions to be determined. When $t\to 0$ we would recover LFT solution (\ref{LFTsolutionAgain}) with its characteristic scale $R_0$. Both the equation of motion and the condition for it to be a vacuum solution are differential equations highly non linear in $r(\psi)$ and its first derivative $r'(\psi)$, but they are linear in the second derivative $r''(\psi)$. Therefore we can combine both equations and establish a non-linear first order differential equation $r(\psi)$ should satisfy to be the inverse of a vacuum solution of the theory:
\begin{align}\label{diffeq1}
0=&\left(16 t \left(t+\psi ^2\right)^3-\left(2 t \psi +\psi ^3\right)^2 r (\psi)^2\right)
-8 \left(t \psi ^3 \left(t+\psi ^2\right) r (\psi)\right) r '(\psi)
\nonumber\\
&+\left(4 \psi ^4 \left(t+\psi ^2\right)^2-\psi ^6 r (\psi)^2\right) r '(\psi)^2-2 \left(\psi ^7 r (\psi)\right) r '(\psi)^3
\end{align}
It should be noted though, that in order to arrive to such a differential equation, a step of squaring an intermediate equation was taken in order to get rid of a square root. This means that any solution of (\ref{diffeq1}) must be checked in the original second order equation of motion to verify no sign ambiguity or constrain remains. We plug the ansatz (\ref{ansatz1}) in (\ref{diffeq1}) and we obtain, order by order, a linear first order differential equation for each function $r_i(\psi)$ which can be easily solved producing one integration constant $c_i$ at each order. Just to illustrate, the first function is
\begin{equation}\label{r1}
r_1(\psi)=
\frac{2}{R_0 \psi }+\frac{c_1(\psi-R_0) }{2 \psi-R_0 }
\end{equation}
with the following ones also being rational functions of $\psi$; {\it c.f.} appendix \ref{app:resummation}. The series goes on indefinitely but the surprise is that the ansatz (\ref{ansatz1}) can be resummed in the compact expression
\begin{empheq}[box=\fbox]{align}\label{Solution1}
|z-z_0|=r(\vp)=
R(t)\sqrt{\tfrac{2\, e^{-\frac{\vp}{2}}}{R(t)\sqrt{\mu }}-1}\ 
%\sqrt{2\mu^{-1/2} R\, e^{-\frac{\vp}{2}}-R^2}\ 
\left(1+2\,t\,\tfrac{\sqrt{\mu}}{R(t)}\,e^{\tfrac{\vp}{2}}
\right),\qquad -\tfrac{1}{4}R(t)^2\leq t\leq 2 R(t)^2
\end{empheq}
if one chooses $R(t)=R_0+R_0 c_1 t+\mathcal{O}(t^2)$ to absorb the series of constants $\{c_i\}_{i\in\mathbb{N}}$ (see appendix \ref{app:resummation} for more details). We shall explain the condition $-\tfrac{1}{4}R(t)^2\leq t\leq 2 R(t)^2$ below.

While we arrived to (\ref{Solution1}) through the ansatz (\ref{ansatz1}) it turns out that solution (\ref{Solution1}) supersedes the ansatz in that it includes solutions with more general $R(t)$ functions which need not have a regular expansion in $t$ close to $t=0$. The role of the function $R(t)$ is that of a $t$-dependent length scale which fixes the maximum of $\vp$ at its center $z_0$ 
\begin{equation}
e^{\vp(z=z_0)}=\frac{4}{\mu R(t)^2}
\end{equation}

If the function $R(t)$ is such that $R(t\!=\!0)=R_0$ then those solutions include the undeformed LFT solution (\ref{LFTsolutionAgain}). A particular one is when the function $R(t)=R_0$; this is, the function $R(t)$ is constant for any value of $t$. We shall call such case the {\it minimal solution} because it is the simplest one which includes the undeformed solution (\ref{LFTsolutionAgain}) at $t=0$.

Let us explain the interval restriction $-\tfrac{1}{4}R(t)^2\leq t\leq 2 R(t)^2$ in detail. Notice that for $|z-z_0|$ to be real or $|z-z_0|^2$ positive, the argument of the square root has to be positive which means $\vp\leq\log(\tfrac{4}{\mu R(t)^2})$. Also in order for $|z-z_0|\geq 0$ the parenthesis on the r.h.s of (\ref{Solution1}) has to be positive which means, combined with the previous bound for $\vp$, that the solution makes sense only for $t\geq -\tfrac{1}{4}R(t)^2$. 

For the other bound recall that while it is true that (\ref{Solution1}) is an exact solution of (\ref{diffeq1}), we warned before that one has to plug this solution back on the second order equation of motion to verify no sign or constraint was missed when squaring the differential equations. Actually, when doing that we find the non-perturbative restriction
\begin{equation}\label{forbidden1}
\text{Sign}\left[R(t)-2 e^{\vp /2} t \sqrt{\mu }+2 e^{\vp } R(t) t \mu \right]-
\text{Sign}\left[R(t)+2\,t\,\sqrt{\mu}\,e^{\tfrac{\vp}{2}} \right]=0
\end{equation}  
The second sign function is always positive from the discussion in the previous paragraph. Therefore, to fulfill (\ref{forbidden1}) we need
\begin{equation}\label{forbidden2}
R(t)-2 e^{\vp /2} t \sqrt{\mu }+2 e^{\vp } R(t) t \mu\geq 0
\end{equation}
A careful analysis of (\ref{forbidden2}) shows that it is always true, for arbitrary values of $\vp\leq\log(\tfrac{4}{\mu R(t)^2})$, as long as $t\leq 2 R(t)^2$. It should be noted that the bound (\ref{forbidden2}) saturates when $r'(\psi)=0$, this is, when it becomes impossible to invert (\ref{Solution1}). This means that the restriction $t\leq 2 R(t)^2$ is not only necessary in order to have a solution of the equation of motion but it also guarantees that $r'(\psi)\geq 0$ and therefore expression (\ref{Solution1}) will have a unique inverse that will allow us to write $\vp(r)=\vp(|z-z_0|)$ unequivocally.

Let us momentarily go back to the {\it minimal solution}. In that case the restriction becomes $t\in\left[-\tfrac{1}{4}R_0^2,2R_0^2\right]$. This means that the minimal generalization of LFT undeformed vacuum has values of $t$ bounded from below and from above in a window that includes $t=0$. It should be stressed though that while the minimal solution has that restriction for $t$, this is not general when we consider arbitrary $R(t)$. In fact, it is possible to choose $R(t)$ such that $t$ is either unbounded from above or unbounded from below or both. Just to show it is possible, we construct the following examples. 

For positive $t$ consider the choice of $R(t)$
\begin{equation}\label{Rexample1}
R(t)=\sqrt{2}\sqrt{t}-R_0
\end{equation}
This $R(t)$ was engineered  such that $R\left(t=2R_0^2\right)=R_0$, this is, it coincides with the minimal solution in its upper bound for $t$. But also notice that the restriction $t\leq 2R(t)^2$, implies for $R(t)$ in (\ref{Rexample1}) that $t\geq 2R_0^2$. This means that for the choice (\ref{Rexample1}) we have that $t$ is unbounded from above. For negative $t$ on the other hand, consider choosing
\begin{equation}\label{Rexample2}
R(t)=4\sqrt{-t}-R_0
\end{equation}
In this case we engineered it such that $R\left(t=-\tfrac{1}{4}R_0^2\right)=R_0$, in other words, such that it coincides with the minimal solution in its lower bound. This time the other restriction $t\geq-\tfrac{1}{4}R(t)^2$, with the choice (\ref{Rexample2}) implies $t\leq-\tfrac{1}{4}R_0^2$. Thus for such choice of $R(t)$, we found a solution in which $t$ is unbounded from below. In fact we can combine the last two choices with the minimal solution to define a continuous $R(t)$ such as
\begin{equation}
R(t)=
\begin{cases}
 4\sqrt{-t}-R_0\qquad  & t <-\tfrac{1}{4}R_0^2\\
 R_0\qquad\qquad\qquad  -\tfrac{1}{4}R_0^2\leq & t \leq 2R_0^2\\
 \sqrt{2}\sqrt{t}-R_0\qquad  & t >2R_0^2 
\end{cases}
\end{equation}
With this choice we can see that $-\tfrac{1}{4}R(t)^2\leq t\leq 2 R(t)^2$ is satisfied always, independently of $t$. Therefore, for this last choice, $t$ is both unbounded from above and from below.

Another particular choice of $R(t)$ is worth mentioning. Consider the scaling symmetry of the equation of motion described in section \ref{ActionAndProperties}, namely $z\to a z$, $t\to |a|^2 t$ and $\vp(z,\bar z)\to\vp(z,\bar z)-\log|a|^2$ ({\it c.f.} equation (\ref{scalingSymmetry})). This symmetry applied to the general solution (\ref{Solution1}) centered at $z_0=0$ maps the solution on itself with $R(t)\to |a|^{-1}R\left(|a|^2 t\right)$. This means that the choice $R(t)=\lambda\sqrt{|t|}$ with $\lambda$ dimensionless is a self-similar solution. Particularly if $\lambda>2$ the solution has no bounds on the allowed values of $t$ though it is ill defined at $t=0$ because $R(0)=0$.

It is clear from these examples that the general solution (\ref{Solution1}) with the restrictions $-\tfrac{1}{4}R(t)^2\leq t\leq 2 R(t)^2$ might imply some kind of bound on $t$ depending on the function $R(t)$. Whether there are or there are not bounds on the parameter $t$ depends on that particular choice and there are infinite choices of $R(t)$ which imply absolutely no bound in $t$. We also stress the fact that it was not necessary to introduce another length scale to overcome the bounds in $t$ as the examples above show: this was possible since a natural length scale is $t$ itself, which has units of $(Length)^2$.

For completeness, we should invert (\ref{Solution1}) to express $\vp(z,\bar z)$. We can reexpress (\ref{Solution1}) as a cubic equation for $\psi=e^{-\tfrac{\vp}{2}}$
\begin{equation}\label{cubic}
0=-4 t^2 R(t)+4 t \left(2 t-R(t)^2\right) \frac{\psi}{\sqrt{\mu}} 
-R(t) \left(r^2-8 t+R(t)^2\right) \left(\frac{\psi}{\sqrt{\mu}}\right)^2+2 R(t)^2 \left(\frac{\psi}{\sqrt{\mu}}\right)^3
\end{equation}
As explained before, there is a unique way of inverting this equation. As long as $-\tfrac{1}{4}R(t)^2\leq t\leq 2 R(t)^2$ there is a unique real solution of the cubic (\ref{cubic}) which is
\begin{equation}\label{Solution1or2Inverted}
%\frac{1}{R(t)\sqrt{\mu}}
e^{-\tfrac{\vp(z,\bar z)}{2}}=\frac{R(t)\sqrt{\mu}}{6}
\left(
\tilde{r}^2-8\tilde{t}+1+\mathfrak{Q}_{+}^{1/3}+s_{\tilde{r},\tilde{t}}|\mathfrak{Q}_{-}|^{1/3}
\right)
\end{equation}
where 
\begin{align}
\mathfrak{Q}_{\pm}=&
\tilde{r}^6+3\tilde{r}^4(1-8\tilde{t})+3\tilde{r}^2(1-4\tilde{t}+40\tilde{t}^2)+(1+4\tilde{t})^3
\nonumber\\
&\pm 12 \tilde{r}\sqrt{3\tilde{t}^2(\tilde{r}^2-r_{+}^2)(\tilde{r}^2-r_{-}^2)},
\quad\mbox{with}\quad r_{\pm}=\left(2\tilde{t}(5+\tilde{t})-1\pm 2\sqrt{\tilde{t}(\tilde{t}-2)^3}
\right)^{1/2}
\end{align}
and
\begin{equation}\label{tildeDefs}
s_{\tilde{r},\tilde{t}}=\mathrm{Sign}\left(\tilde{r}^4+\tilde{r}^2(2-16\tilde{t})+(1+4\tilde{t})^2\right),\qquad
\tilde{r}=\frac{|z-z_0|}{R(t)},\qquad \tilde{t}=\frac{t}{R(t)^2}
\end{equation}
It is curious that $\vp=-2\log|z|^2+\mathcal{O}(1)$ when $|z|\to\infty$ which is the same behavior as the undeformed solution.

Notice that in order to write the cubic (\ref{cubic}) we had to take a square of the implicit solution (\ref{Solution1}) which makes one wonder what happens with the reversed sign solution
\begin{equation}\label{Solution2}
|z-z_0|=r(\vp)=
R(t)\sqrt{\tfrac{2\, e^{-\frac{\vp}{2}}}{R(t)\sqrt{\mu }}-1}\ 
%\sqrt{2\mu^{-1/2} R\, e^{-\frac{\vp}{2}}-R^2}\ 
\left(-1-2\,t\,\tfrac{\sqrt{\mu}}{R(t)}\,e^{\tfrac{\vp}{2}}
\right),\qquad t\leq -\tfrac{1}{4}R(t)^2
\end{equation}
This solution candidate makes sense only for $t\leq -\tfrac{1}{4}R(t)^2$ such that $|z-z_0|>0$. The problem with this solution is that it is incomplete; let us see how. If one tries to invert (\ref{Solution2}) in order to have a univalued function $\vp(z,\bar z)$ one finds there are two branches with $\psi=e^{-\vp/2}$ real and positive. The turning point of those two branches is $\vp=\vp_0$ with $e^{-\vp_0/2}=\sqrt{\mu}R(t)(\tilde t+\sqrt{\tilde t(\tilde t-2)})$ and $\tilde t$ as in (\ref{tildeDefs}). Plugging (\ref{Solution2}) into the equation of motion one finds the constraint $\vp>\vp_0$ which selects the upper branch. But this at the same time means that (\ref{Solution2}) is only defined for values of $0\leq|z-z_0|<r_0$ with $r_0$ given by
\begin{equation}
r_0=r(\vp_0)=R(t)\left(2\tilde{t}^2+10\tilde t-2+2\sqrt{\tilde t(\tilde t-2)}\right)^{1/2}
\end{equation} 
In other words, with solution (\ref{Solution2}) we do not find a function $\vp(z,\bar z)$ defined for the whole complex plane but only for the disk $0\leq|z-z_0|<r_0$.

We end up this section by showing that the solution we called minimal, this is (\ref{Solution1or2Inverted}) with the simplest choice $R(t)=R_0$ corresponds to the vacuum of the deformed theory if one insists on interpreting the $T\bar T$ deformation geometrically as a field dependent coordinate transformation \cite{Conti:2018tca,Dubovsky:2017cnj}. Solving the differential equations for the change of variables\footnote{See formula 5.1 in \cite{Conti:2018tca}} applied to Liouville undeformed solution (\ref{LFTsolutionAgain}) one finds
\begin{equation}\label{systemVariables}
z-z_0=w-w_0+\frac{4 t (w-w_0)}{R_0^2+|w-w_0|^2},\qquad
\bar z-\bar{z}_0=\bar w-\bar{w}_0+\frac{4 t (\bar w-\bar{w}_0)}{R_0^2+|w-w_0|^2}
\end{equation}
where $(w,\bar w)$ and $(z,\bar z)$ are the undeformed and the deformed coordinates respectively. We have checked that inverting (\ref{systemVariables}) and plugging $(w(z,\bar z),\bar w(z,\bar z))$ in the undeformed solution one obtains the minimal solution, this is (\ref{Solution1or2Inverted}) with $R(t)=R_0$. Recall the minimal solution had the restriction $-\tfrac{1}{4}R_0^2<t<2R_0^2$, {\it i.e.} $t$ is bounded both from below and from above. One can conclude from this that the $T\bar T$ deformation of LFT, interpreted as a field dependent coordinate change, can not be extended to arbitrary values of $t$ but only to an interval around $t=0$. Since our whole analysis of this section also started with expansions of solutions around $t=0$ and even if we were able to generalize those expansions by insisting on an arbitrary $R(t)$, it remains an open question whether the solutions we found are the only vacuum solutions of the $T\bar T$ deformed theory.

%We end up this section by making the following consistency check. Consider again the original undeformed LFT solution (\ref{LFTsolutionAgain}). It is easy to show that it satisfies $\dvp\dbvp=\tfrac{2\sqrt{\mu} e^{\vp/2}}{R_0}-\mu e^{\vp}$. One can take this last expression and apply the field-dependent change of variables of \cite{Conti:2018tca,Dubovsky:2017cnj} to obtain the following non-linear first order differential equation
%\begin{equation}\label{ChangeEq}
%\frac{2\,\sqrt{\mu}\, e^{\vp(r)/2}}{R_0}-\mu\,e^{\vp(r)}=\frac
%{\left(1-\sqrt{1+t\,\vp'(r)^2\,(1+\mu\, t\, e^\vp)}\right)^2}
%{t^2\,\vp'(r)^2}
%\end{equation}
%We have checked that what we called the {\it minimal solution} is one of the two solutions of (\ref{ChangeEq}) (the other one being non-physical).

\section{Discussion}\label{sec:discussion}

When we presented classical Liouville field theory at the beginning of this work we said its local properties can be derived from the Lagrangian $\mathcal{L}^{(0)}=\dvp\dbvp-\mu e^{\vp}$ which has to be supplemented with the condition $\vp(z,\bar z)=-2\log(z\bar z)+\mathcal{O}(1)$ for $|z|\to\infty$ to have the theory globally defined on the Riemann sphere. Some authors instead (see \cite{Zamolodchikov:1995aa} for a discussion on this issue) conventionally add an extra term in the Lagrangian
\begin{equation}\label{modLFT}
\mathcal{L}^{(0)}=\frac{1}{4}\hat{g}^{ab}\partial_{a}\vp\partial_{b}\vp-\mu e^{\vp}+\tfrac{1}{2}\hat{R}\vp
\end{equation}
where $\hat g$ and $\hat{R}$ refer to a background metric and corresponding curvature. This is, even if one studies the theory in flat space and $\hat{R}=0$ in every local expression, the additional term of (\ref{modLFT}) accounts for the coupling of Liouville field with the curvature at $|z|\to\infty$ to enforce the boundary condition $\vp(z,\bar z)=-2\log(z\bar z)+\mathcal{O}(1)$ for $|z|\to\infty$ through this so called background charge. One of the advantages of such formulation is that if instead of using the canonical definition of the stress tenor one uses Hilbert definition by the background metric $T^{(h)}_{\mu\nu}=\tfrac{2}{\sqrt{\hat g}}\tfrac{\delta(\sqrt{\hat g}\mathcal{L}^{(0)})}{\delta\hat{g}^{\mu\nu}}$ an holomorphic stress tensor component $\t^{(h)}=-(\dvp)^2+2\partial^2\vp$ with trace $\Theta^{(h)}=0$ is immediately obtained without the need to add ad-hoc modifications to the canonical stress tensor. One may wonder whether another type of a $T\bar T$ deformation of LFT could be studied if the irrelevant flow is driven by the Hilbert stress tensor, this is if $\partial_t\mathcal{L}^{(t)}=\det(T^{(h)}_{\mu\nu})$. The authors of \cite{Bonelli:2018kik} have already studied the $T\bar T$ flow of the addition of linear dilaton coupling such as the one we added in (\ref{modLFT}), at least perturbatively  in $t$. The conclusion for the first orders in $t$ is that one obtains a theory with infinite higher derivative terms which seems intractable non-perturbatively in the $T\bar T$ flow. Thus, as far as our understanding goes, the $T\bar T$ deformation driven by the canonical stress tensor we studied in this work seems the only tractable non-perturbative problem for $T\bar T$ deformations of classical LFT.

In section 3 of this work we were able to construct an infinite set of non trivial higher conserved currents of the $T\bar T$ deformation of classical LFT. Besides the current pairs we obtained $\{\t_{2n},\Theta_{2n\!-\!2}\}$ (and their opposite spin counterparts) other higher derivative currents such as the pair $\{\t_{2n\!+\!\textbf{s}},\Theta_{2n\!-\!2\!+\!\textbf{s}}\}=\{\partial^{\textbf{s}}\t_{2n},\partial^{\textbf{s}}\Theta_{2n\!-\!2}\}$ are trivially conserved. The existence of these infinite towers of conserved currents is an indication that the theory is classically integrable. In fact, using these higher conserved currents we were able to obtain a first order differential equation for vacuum solutions.  For future research it would be interesting to construct a Lax-pair formulation of the integrability problem for this $T\bar T$ deformed theory such as the one constructed for $T\bar T$ deformed $\sin(h)$-Gordon theory in \cite{Conti:2018jho}. The vacua of $T\bar T$ deformed Liouville, which we obtained using some educated guesses and variable changes, might be derivable from symmetry and integrability properties if those structures actually exist. The vacua of classical LFT is the trivial solution to the sphere uniformization through the connection of classical LFT and the Riemann-Hilbert problem. Non trivial stress tensors in classical LFT correspond to the punctured sphere with parabollic or hyperbolic singularities. It would be interesting to understand the connection of the $T\bar T$-deformed version of classical LFT we studied in this work with uniformization problems. Additionally, it is well known that classical LFT satisfies an infinite tower of higher equations of motion \cite{Zamolodchikov:2003yb}, a $T\bar T$ version of which would be desirable.

\acknowledgments

We would like to thank Gaston Giribet and Marco S. Bianchi for relevant comments on a draft version of this work.

%\paragraph{Note added.} ...

%\newpage

\appendix
\section{Derivation of the Lagrangian}\label{app:Lagrangian}

With the use of recurrence (\ref{RecurrenceLagrangian}) and starting with $\mathcal{L}_0=X+V$ the first few terms are
\begin{align}
&\mathcal{L}_1=-X^2+V^2,\quad \mathcal{L}_2=2X^3+VX^2+V^3,
\quad \mathcal{L}_3=-5X^4-4VX^3+V^4,\nonumber\\
&\mathcal{L}_4=14X^5+15VX^4+2V^2X^3+V^5,\quad
\mathcal{L}_5=-42 X^6-56VX^5-15V^2X^4+V^6
\end{align}
which lead to the pattern
\begin{align}
&\mathcal{L}_n=V^{n+1}+(-1)^n\sum_{k=0}^{[\tfrac{n}{2}]}
\frac{(2n-2k)!\, X^{n-k+1}\,V^k}{k!\,\,(n-k+1)!\,(n-2k)!}\quad\mbox{and} \label{app:ExactSum01.1}\\
&\mathcal{L}_{n,X}=(-1)^n\sum_{k=0}^{[\tfrac{n}{2}]}
\frac{(2n-2k)!\, X^{n-k}\,V^k}{k!\,\,(n-k)!\,(n-2k)!}\label{app:ExactSum01.2}
\end{align}
From (\ref{app:ExactSum01.1}) we see $\mathcal{L}_n(X=0,V)=V^{n+1}$ such that $\mathcal{L}^{(t)}(X=0,V)=\tfrac{V}{1-t\,V}$. We focus on (\ref{app:ExactSum01.2}). It is possible to extend the sum in $k$ up to infinity by exchanging the factorials by appropriate $\Gamma$ functions $(..)!\to\Gamma(..+1)$ and observing that the added new terms for $k>\left[\tfrac{n}{2}\right]$ are all zero. Thus we can write
\begin{equation}
\mathcal{L}^{(t)}_X=\sum_{n=0}^{\infty}\sum_{k=0}^\infty\frac{\Gamma(2n-2k+1)(-tX)^n(V/X)^k}
{k!\Gamma(n-k+1)\Gamma(n-2k+1)}
\end{equation}
The idea is now to exchange the order of the series and observe that the first $2k-1$ values for the sum in $n$ vanish. We get
\begin{equation}
\mathcal{L}^{(t)}_X=\sum_{k=0}^{\infty}\sum_{n=2k}^\infty\frac{\Gamma(2n-2k+1)(-tX)^n(V/X)^k}
{k!\,\Gamma(n-k+1)\Gamma(n-2k+1)}
\end{equation} 
Shifting the $n$ series to start from zero and using properties of the $\Gamma$ function we simplify the expression
\begin{equation}\label{app:ExactSum02}
\mathcal{L}^{(t)}_X=\frac{1}{\sqrt{\pi}}\sum_{k=0}^{\infty}\sum_{n=0}^\infty\frac{\Gamma(\tfrac{1}{2}+n+k)}
{k!\,n!}(-4tX)^n(4t^2XV)^k
\end{equation}
And this is the iterated double binomial series we mention in (\ref{ExactSum02}).

\section{Conservation of currents}\label{app:currents}

With the case $n=1$ proven in (\ref{proof0}) we can show the conservation (\ref{DefConservation}) for $n$ inductively. Assuming (\ref{DefConservation}) holds for $n$ we consider the difference $\bar\d\t_{2n+2}-\d\theta_{2n}$
\begin{align}
&\bar\d\t_{2n+2}-\d\theta_{2n}=\d\left(\frac{\Omega(\dvp)^2}{4t X^2}\Theta_0\Theta_{2n-2}\right)
-\bar\d\left(\frac{\Omega(1-tV)}{(1+\Omega)^2}\t_2\t_{2n}\right)\label{app:proof1}\\
=&\d\left(\frac{\Omega(\dvp)^2}{4t X^2}\Theta_0\right)\Theta_{2n-2}
-\bar\d\left(\frac{\Omega(1-tV)}{(1+\Omega)^2}\t_2\right)\t_{2n}\label{app:proof2}\\
=&\Theta_{2n\!-\!2}\left[
\d\left(\frac{\Omega(\dvp)^2}{4t X^2}\Theta_0\right)
-\frac{4t(1-tV)(\dvp)^2}{(1-\Omega)^2}\bar\d\left(\frac{\Omega(1-tV)}{(1+\Omega)^2}\t_2\right)
\right]\label{app:proof3}\\
=&\Theta_{2n\!-\!2}\left[
\d\left(\frac{\Omega(\dvp)^2}{4t X^2}\right)\Theta_0
-\frac{4t(1-tV)(\dvp)^2}{(1-\Omega)^2}\bar\d\left(\frac{\Omega(1-tV)}{(1+\Omega)^2}\right)\t_2
\right]\label{app:proof4}\\
=&4t(1-tV)\Theta_{2n\!-\!2}\t_2\left[
\frac{(\dbvp)^2}{(1+\Omega)^2}\d\left(\frac{\Omega(\dvp)^2}{4t X^2}\right)-
\frac{(\dvp)^2}{(1-\Omega)^2}\bar\d\left(\frac{\Omega(1-tV)}{(1+\Omega)^2}\right)
\right]=0\label{app:proof5}
\end{align}
In equality (\ref{app:proof1}) we used the recursive property (\ref{RecursionTauTheta1}) and to go from (\ref{app:proof1}) to (\ref{app:proof2}) we used the inductive hypothesis. From (\ref{app:proof2}) to (\ref{app:proof3}) we used the second relation in (\ref{RecursionTauTheta2}) for $\t_{2n}$ and from there to (\ref{app:proof4}) we used the conservation of $\t_2$ and $\Theta_0$ shown in (\ref{proof0}). Using the first relation of (\ref{RecursionTauTheta2}) for $n=1$ we arrive to (\ref{app:proof5}) which is the expression mentioned in the main text (\ref{proof2}).

\section{Resummation}\label{app:resummation}

With the ansatz
\begin{equation}\label{app:ansatz}
|z-z_0|=r(\psi)=R_0\sqrt{\tfrac{2\, \psi}{R_0}-1}\ \Big(1 
+t\,r_1(\psi)+t^2\,r_2(\psi)+...\Big)
\end{equation}
and the differential equation for the inverse of vacuum solutions
\begin{align}\label{app:diffeq}
0=&\left(16 t \left(t+\psi ^2\right)^3-\left(2 t \psi +\psi ^3\right)^2 r (\psi)^2\right)
-8 \left(t \psi ^3 \left(t+\psi ^2\right) r (\psi)\right) r '(\psi)
\nonumber\\
&+\left(4 \psi ^4 \left(t+\psi ^2\right)^2-\psi ^6 r (\psi)^2\right) r '(\psi)^2-2 \left(\psi ^7 r (\psi)\right) r '(\psi)^3
\end{align}
one obtains the functions
\begin{equation}\label{app:r1}
r_1(\psi)=
\frac{2}{R_0 \psi }+\frac{c_1(\psi-R_0) }{2 \psi-R_0 }
\end{equation}
\begin{align}\label{app:r2}
r_2(\psi)=&
\frac{\left.8 R_0 c_1+R_0^3 \left(c_1^2+4 c_2\right)\right)-\left(c_1 \left(16+3 R_0^2 c_1\right)+12 R_0^2 c_2\right) \psi +8 R_0 c_2 \psi ^2}
{4 R_0 (2 \psi-R_0 )^2}
\end{align}
\begin{align}\label{app:r3}
r_3(\psi)=& \frac{1}{4 R_0 (2 \psi-R_0 )^3}\Big(
16 R_0 c_3 \psi ^3\!-\!\left(\!-\!16 c_1^2\!+\!R_0^2 c_1^3\!+\!12 R_0^2 c_1 c_2\!+\!32 \left(c_2\!+\!R_0^2 c_3\right)\right) \psi ^2
\nonumber\\
&\!+\!2 R_0 \left(\!-\!10 c_1^2\!+\!16 c_2\!+\!5 R_0^2 c_1 c_2\!+\!10 R_0^2 c_3\right) \psi 
\!-\! 2 R_0^2 \left(-3 c_1^2\!+\!4 c_2\!+\!R_0^2 c_1 c_2\!+\!2 R_0^2 c_3\right)
\Big)
\end{align}
and so on. This infinite series can be resummed in the solution
\begin{align}
|z-z_0|=r(\vp)=
R(t)\sqrt{\tfrac{2\, e^{-\frac{\vp}{2}}}{R(t)\sqrt{\mu }}-1}\ 
%\sqrt{2\mu^{-1/2} R\, e^{-\frac{\vp}{2}}-R^2}\ 
\left(1+2\,t\,\tfrac{\sqrt{\mu}}{R(t)}\,e^{\tfrac{\vp}{2}}
\right),\qquad -\tfrac{1}{4}R(t)^2\leq t\leq 2 R(t)^2
\end{align}
with
\begin{equation}
R(t)=R_0+R_0 c_1 t+\frac{1}{4} R_0 \left(c_1^2+4 c_2\right) t^2+R_0\frac{1}{2} \left( c_1 c_2+2c_3\right) t^3+\mathcal{O}(t^4)
\end{equation}

% The bibliography will probably be heavily edited during typesetting.
% We'll parse it and, using the arxiv number or the journal data, will
% query inspire, trying to verify the data (this will probalby spot
% eventual typos) and retrive the document DOI and eventual errata.
% We however suggest to always provide author, title and journal data:
% in short all the informations that clearly identify a document.

\end{document}